\documentclass[aps,prb,superscriptaddress,showpacs]{revtex4}
\usepackage{graphicx}

\begin{document}
\title{Doping effects on the electronic and structural properties of 
CoO$_2$: An LSDA+U study}

\author{Peihong Zhang}
\author{Weidong Luo}
\affiliation{
Department of Physics, University of California at Berkeley, Berkeley, 
California 94720}
\affiliation{
Materials Sciences Division, Lawrence Berkeley National Laboratory, Berkeley, 
California 94720}
\author{Vincent H. Crespi}
\affiliation{
Department of Physics and Materials Research Institute, The Pennsylvania State University,
University Park, PA 16802}
\author{Marvin L. Cohen}
\author{Steven G. Louie}
\affiliation{
Department of Physics, University of California at Berkeley, Berkeley, 
California 94720}
\affiliation{
Materials Sciences Division, Lawrence Berkeley National Laboratory, Berkeley, 
California 94720}

\date{\today}

\begin{abstract}
A systematic LSDA+U study of doping effects on the electronic and
structural properties of single layer CoO$_2$ is presented. Undoped 
CoO$_2$ is a charge transfer insulator within LSDA+U and
a metal with a high density of states (DOS) at the Fermi level within LSDA.
(CoO$_2$)$^{1.0-}$, on the other hand, is a band insulator with a gap of 2.2 eV.
Systems with fractional doping are metals if no charge orderings are present.
Due to the strong interaction between the doped electron and other
correlated Co $d$ electrons, the calculated electronic structure of (CoO$_2$)$^{x-}$ 
depends sensitively on the doping level $x$. Zone center optical phonon energies 
are calculated under the frozen phonon approximation and are in good agreement 
with measured values. Softening of the $E_g$ phonon at doping $x\sim 0.25$
seems to indicate a strong electron-phonon coupling in this system.
Possible intemediate spin states of Co ions, Na ordering, 
as well as magnetic and charge orderings in this system are also discussed. 
\end{abstract}

\maketitle

\section{Introduction}

The recent discovery of superconductivity in hydrated Na$_x$CoO$_2$\cite{supercond} 
has generated renewed interest in this technologically important material. Na$_x$CoO$_2$ 
has been known for several years as a potential thermoelectric material which exhibits 
an unexpectedly large thermoelectric power and at the same time a low 
resistivity\cite{thermopower}. Although the origin of the large thermopower remains a 
subject of active investigation, strong correlations between Co $d$ electrons and spin 
entropy are believed to play a critical role\cite{correlation1,correlation2,spin_entropy}. 
Beside its unusual thermoelectric properties, Na$_x$CoO$_2$ ($x$ in the range 0.5 $\sim$ 0.75) 
is also known for having a Curie-Weiss type of susceptibility instead of a Pauli 
paramagnetic behavior\cite{metallic,susceptibility,magnetic_transition1}, which would be 
more compatible with its metallic conduction\cite{metallic}. 
Although there have been reports of a weak magnetic ordering transition 
in Na$_{0.75}$CoO$_2$ at $T_m=$ 22 K\cite{magnetic_transition1,magnetic_transition2}, 
no such transition has been observed down to 2 K for systems with lower Na contents. 
Compared to the vast experimental work that has been done on this material, 
theoretical study seems to have lagged behind. 

First principle calculations of electronic and magnetic properties of strongly correlated 
systems such as Na$_x$CoO$_2$ have always been a challenge. Although the local spin 
density approximation (LSDA) to the density functional theory (DFT) has been applied to 
various systems with great success, it is well known that the LSDA fails in many aspects 
when applied to late transition metal oxides in which strong correlations between $d$ 
electrons play an important role.
For example, LSDA fails 
to reproduce the insulating, antiferromagnetic (AFM) ground state for several transition 
metal oxides, including the parent materials of high transition temperature (T$_c$) 
superconductors. Not surprisingly, there have been several attempts to improve the L(S)DA 
to take into account (at least partially) the strong electron-electron interactions
in first-principle calculations. One of the simplest, yet very successful schemes, was 
proposed by Anisimov et al.\cite{LDA+U1,LDA+U2,LDA+U3,LDA+U4}:
the LDA+U method. In this paper, we present a systematic LSDA+U study of doping effects 
on the electronic and structural properties of Na$_x$CoO$_2$ ($0 \leq x \leq 1$) using 
a recently implemented rotationally invariant LSDA+U method\cite{LDA+U4} within the 
pseudopotential plane-wave formalism\cite{pseudo}.

\section{The LSDA+U method}

It is now well established that the failure of LSDA, when applied 
to late transition metal oxides, is largely due to an insufficient 
treatment of on-site Coulomb interactions between the rather 
localized $d$ electrons. LSDA attempts to account for the Coulomb 
interaction via an averaged potential depending only on local spin densities. 
Consequently, magnetic moment formation is driven mainly by the 
spin polarization energy within LSDA. Orbital polarizations, on the
other hand, play less important roles in LSDA and the occupation of localized 
orbitals does not depend sensitively on their orientation (symmetry). 
For highly localized electrons, however, the Coulomb interaction should be applied to 
the localized orbital as a whole and is better described by the Hubbard 
or Hartree-Fock (HF) type of theory. The formation of a local moment is 
therefore a result of both spin and orbital polarizations. Unfortunately, 
although the HF aproximation and other correlated quantum chemistry
methods have been applied to atomic and molecular 
systems with tremendous success, their application to solids has been 
limited. Screening effects, which are relatively weak in atoms 
and molecules, are usually important in solids and cannot be easily 
included within the HF theory. Higher level quantum chemistry
calculations, on the other hand, are computationally too expensive for 
most solids.

The LSDA+U method attempts to incorporate the orbital 
specific screened Coulomb interaction while retaining
the simplicity of LSDA. In LSDA+U, the energy 
functional consists of three contributions\cite{LDA+U1,LDA+U2}:
\begin{equation}
E^{\mathrm{LSDA+U}}[\rho^\sigma({\bf r}),\{{\bf n}^\sigma\}]=
E^{\mathrm{LSDA}}[\rho^\sigma({\bf r})]+E^U[\{{\bf n}^\sigma\}]-E^{dc}[\{{\bf n}^\sigma\}],
\label{ELSDA+U}
\end{equation}

\noindent where $E^{\mathrm{LSDA}}$ is the usual LSDA energy functional for spin 
densities $\rho^\sigma$ ($\sigma=\uparrow,\downarrow$), $E^U$ is a Hubbard or 
HF type of interaction arising from localized electrons (described by 
orbital occupation density matrices ${\bf n}^\sigma$) and $E^{dc}$ is a 
``double-counting'' term to be defined later. In the rotationally invariant 
LSDA+U method\cite{LDA+U4}, $E^U$ takes the familiar HF form:
\begin{equation}
E^U=\frac{1}{2}\sum_{\{m\},\{\sigma\}}(\langle m_1,m_2|V^{ee}|m_3,m_4\rangle
-\delta_{\sigma,\sigma^\prime}\langle m_1,m_2|V^{ee}|m_4,m_3\rangle)
n^\sigma_{m_1,m_3}n^{\sigma^\prime}_{m_2,m_4},
\label{U}
\end{equation}

\noindent where the matrix elements of the screened electron-electron 
interaction $V^{ee}$ can be expressed approximately as a sum of Slater integrals 
$F^k$:

\begin{equation}
\langle m_1,m_2|V^{ee}|m_3,m_4\rangle\approx\sum_{k=0,2}^{2l} a_k(m_1,m_3,m_2,m_4)F^k,
\label{Vee}
\end{equation}

\begin{equation}
F^k \approx \frac{1}{\epsilon}\int{\frac{r^k_{<}}{r^{k+1}_{>}}R^2_l(r_1)R^2_l(r_2)dr_1dr_2},
\label{Fk}
\end{equation}

\begin{equation}
a_k(m_1,m_3,m_2,m_4)=\frac{4\pi}{2k+1}\sum_{q=-k}^k \langle Y_{lm_1}|Y_{kq}|Y_{lm_3}\rangle.
\end{equation}

\noindent Here $\epsilon$ is the dielectric constant of the system,
$r_<$ and $r_>$ are the smallest and largest values of $r_1$ and $r_2$,
and $R_l$ is the radial wavefunction of the localized electron.
It should be pointed out that the Slater integrals are not well defined in 
solids and expression (\ref{Vee}) is only an approximation. 
For $d$ electrons ($l=2$), three Slater integrals, $F^0$, $F^2$ and
$F^4$, are needed. 
The Slater integrals 
relate to the familiar Coulomb (or Hubbard) $U$ and exchange $J$ parameters a
as $U=F^0$ and $J=(F^2+F^4)/14$. A further simplication can be achieved 
by the observation that $F^4/F^2\approx 0.625$ for most $d$-electron 
systems\cite{LDA+U2,Groot90}. The Slater integrals $F^k$'s (or equivalently, $U$ and $J$) 
are fixed parameters in our calculations. In principle, however, one could 
calculate these parameters self-consistently. The double counting term
\begin{equation}
E^{dc}[\{n^\sigma\}]=U\frac{n(n-1)}{2}
-J[\frac{n^\uparrow(n^\uparrow-1)}{2}+\frac{n^\downarrow(n^\downarrow-1)}{2}]
\label{DC}
\end{equation}

\noindent is the averaged electron-electron interaction already included 
in LSDA, assuming that LSDA gives the overall Coulomb and exchange energies
correctly. 
The double counting term
is not uniquely defined and there have been some discussions in
the literature concerning other possible forms 
and their effects on the calculated properties\cite{DC}. 
In the above expressions, $n^\sigma=$Tr(${\bf n}^\sigma$) and
$n=n^\uparrow+n^\downarrow$, while the density matrix
${\bf n}^\sigma$ for localized (e.g., $d$ or $f$) orbitals
remains to be defined.
Identifying localized orbitals is trivial in 
computational methods using atomic basis sets such as the linear muffin-tin orbital 
method (LMTO). In the pseusopotential plane-wave method, this is less
obvious and can be done by projecting the wavefunctions $\Psi^\sigma_{nk}$
onto pseudoatomic orbitals $R_l(r)$ calculated with an appropriate atomic configuration
($3d^64s^0$ for Co in this work):
\begin{equation}
n^\sigma_{m_1,m_2}=\sum_{n,k}\langle m_1|\Psi^\sigma_{nk}
\rangle\langle\Psi^\sigma_{nk} |m_2\rangle.
\label{density_matrix}
\end{equation}

\noindent We have used the abbreviation $|m\rangle\equiv|R_lY_{lm}\rangle$.
Note that proper symmetrizations of the density matrix are  
needed if the above summation is carried out in the irreducible Brillouin zone (BZ).
Diagonalization of the density matrix then gives the 
symmetry-adapted local orbitals and their occupation numbers. 

Applying the variational principle to the energy functionals
defined in Eqs. (\ref{ELSDA+U}), (\ref{U}) and (\ref{DC}), we have, in addition 
to the usual one-electron LSDA Hamiltonian, an orbital-dependent correction term
\begin{equation}
\delta {\bf V}^\sigma=\sum_{m_1,m_2}|m_1\rangle\delta V_{m_1,m_2}^\sigma\langle m_2|,
\end{equation}

\noindent where the matrix elements
\begin{equation}
\delta V_{m_1,m_2}^\sigma=\sum_{m_3,m_4,\sigma^\prime}(\langle m_1,m_3|V^{ee}|m_2,m_4\rangle-
\delta_{\sigma\sigma^\prime}\langle m_1,m_3|V^{ee}|m_4,m_2\rangle)n_{m_3,m_4}^{\sigma^\prime}
-\delta_{m_1,m_2}[U(n-\frac{1}{2})+J(n^\sigma-\frac{1}{2})].
\label{deltaV}
\end{equation}

\noindent The resulting one-electron problem 
\begin{equation}
(H^\sigma_{\mathrm{LSDA}}+\delta{\bf V}^\sigma)|\Psi_{nk}^\sigma\rangle=E_{nk}^{\sigma}
|\Psi_{nk}^\sigma\rangle
\label{LSDA+U}
\end{equation}

\noindent can be solved self-consistently. Due to the presence of the orbital-dependent 
potential ${\bf V}^\sigma$, it is more convenient to solve Eq. (\ref{LSDA+U}) in two 
steps\cite{Shick99}. First, we solve an auxiliary LSDA problem
\begin{equation}
H^\sigma_{\mathrm{LSDA}}|\Phi_{nk}^\sigma\rangle=\varepsilon_{nk}^{\sigma}|\Phi_{nk}^\sigma\rangle
\end{equation}

\noindent to obtain an orthogonal basis \{$\Phi_{nk}^\sigma$\} and the corresponding
eigenvalues \{$\varepsilon_{nk}^\sigma$\}. Note that the one-electron
Hamiltonian $H^\sigma_{\mathrm{LSDA}}$ is constructed using the electron density determined
by minimizing the LSDA+U (not the LSDA) energy functional. We then construct the full 
LSDA+U Hamiltonian matrix
\begin{equation}
[H^\sigma_{\mathrm{LSDA+U}}(\vec{k})]_{n,n^\prime}=\delta_{n,n^\prime}\varepsilon^{\sigma}_{nk}+
\langle\Phi_{nk}^\sigma|\delta{\bf V}^\sigma|\Phi_{n^\prime k}\rangle
\end{equation}

\noindent on the subspace of interest (e.g., Co $d$ orbitals) 
and diagonalize it to obtain the LSDA+U wavefunctions $\Psi_{nk}^\sigma$
and band energies $E^\sigma_{nk}$. The occupation matrix ${\bf n}^\sigma$ and 
charge density $\rho^\sigma$ are then constructed for the next iteration until the
self-consistency is achieved, while fixing the parameters $U$ and $J$.

\section{Computational details}

\begin{figure}[h]
\includegraphics[width=8cm]{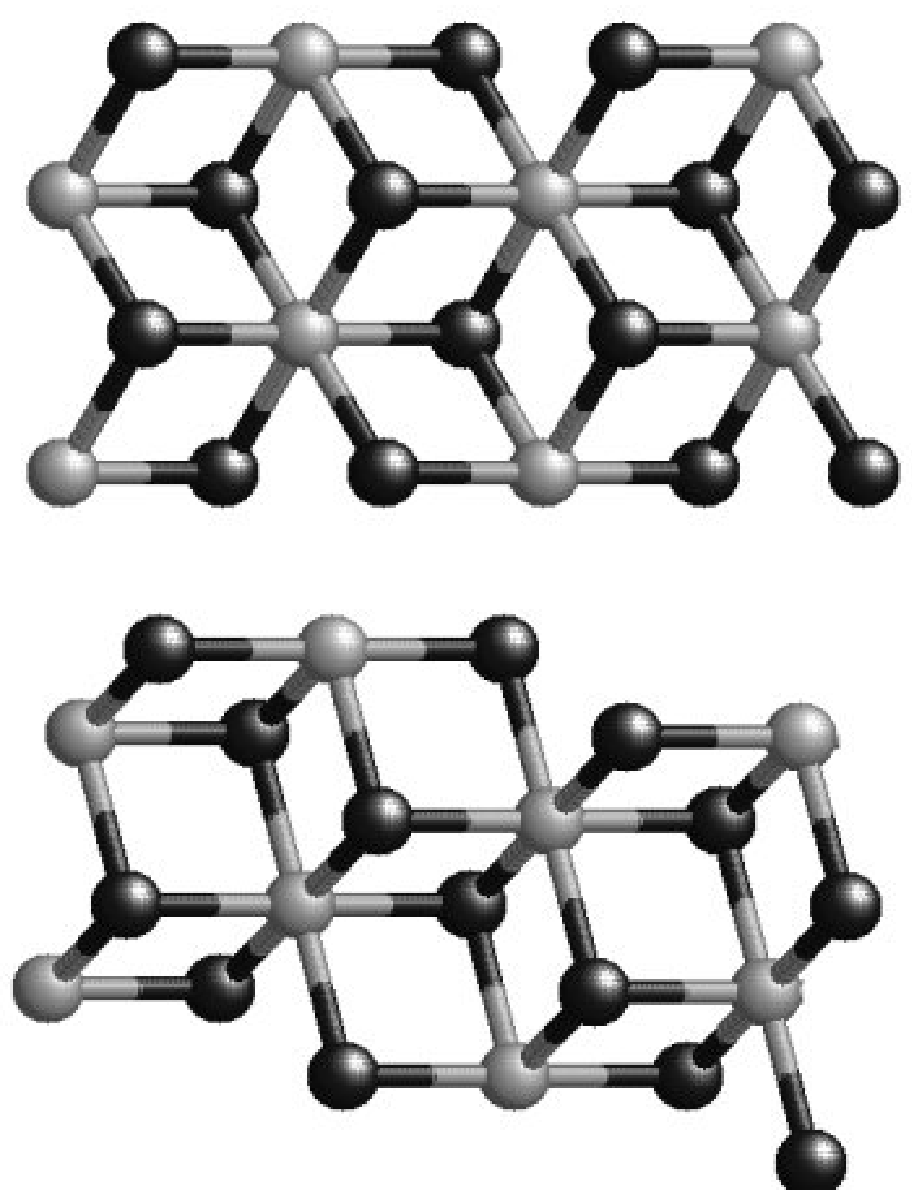}
\caption{Orthographic (upper) and perspective (lower) views of single layer CoO$_2$. Dark balls
are oxygen and gray ones are Co.}
\label{model}
\end{figure}

Na$_x$CoO$_2$ assumes a layered structure in which CoO$_2$ and Na 
layers alternate along the $c$ axis. The electronically active CoO$_2$ 
layer consists of edge sharing CoO$_6$ octahedra with magnetic (Co)
ions forming a frustrated triangular lattice (see Fig.\ \ref{model}). 
The oxygen octahedra 
are distorted considerably - compressed along the body-diagonal direction
of the embedding rocksalt structure and stretched in the perpendicular 
plane. The distortion presumably depends on the doping level (as will
be shown later). This active CoO$_2$ layer is
believed to be responsible for various abnormal electronic, magnetic 
and transport properties of the system and is the focus of the present study. 
The sodium layer is disordered, with Na ions distributed among 
two distinct, partially occupied sites (Wyckoff indices 2$b$ and 2$d$). 
In order to avoid inconvenient (i.e., large) unit cells for systems with
fractional doping, the effects of Na is modeled by corresponding electron
doping into the CoO$_2$ layer in our calculations. The excess electrons 
are then balanced by a uniform postive background. In real systems, 
the presence of Na potentials and small strains associated with 
them, as well as interlayer interactions, might have some additional effects on the
electronic properties. We believe, however, our model captures the 
essential physics and the effects of Na ions are minor if not negligible, 
as will be discussed later. (Alternatively, one may employ the virtual crystal 
technique to overcome the large unit cell problem.) Our treatment 
might become even more exact in the case of hydrated compounds since 
water molecules are likely to screen out the Na potentials. Of cource, 
the interaction between H$_2$O and CoO$_2$ layers is another issue that
deserves further investigations. In our calculations for single layer 
CoO$_2$, we fix the in-plane lattice constants 
$a = b = 2.823$\AA\cite{supercond}, regardless of the doping level. 
Small variation in lattice constants should have negligible 
effects on our results. The separation bewteen layers is set at 6.5 \AA~to 
ensure no significant interlayer interactions.
Therefore, the only structural parameter allowed to relax is the oxygen 
$z$ coordinate, which turns out to be rather sensitive to the doping 
level, as will be discussed later.

The above assumptions significantly simplify our calculations. There are, 
however, other difficulties arised in studying correlated systems
due to the existence of several competing charge and/or magnetic orderings.
Since the magnetic and/or charge ordering energies are usually very small,
it is sometimes difficult to distinguish between different ordering states 
based only on their energy difference. This is particularly true in 
magnetically frustrated systems such as CoO$_2$, as is evidenced
by experiments where no long range magnetic ordering is observed
for Na$_x$CoO$_2$ except for $x\sim 0.75$\cite{magnetic_transition1,magnetic_transition2}.
Nevertheless, it was suggested that a short range ferromagnetic (FM) ordering might 
be preferred in these systems\cite{NaCo2O4}. Therefore, we will primary 
concentrate on the ferromagnetic (paramegtic for $x=1$) phase in this paper. 
Different orderings will be discussed briefly.

We employ the LSDA+U method as described in section II to study doping
effects on the electronic, structural and magnetic properties of
single layer (CoO$_2$)$^{x-}$ ($0\leq x\leq 1$). The $k$-point set is
generated by the Monkhorst-Pack scheme\cite{kpoint} with a density of 
12$\times$12$\times$2. The plane-wave energy cutoff is set at 250 Ry to 
ensure the convergence of the calculations. Such a high plane-wave energy  
cutoff is necessary for systems containing very localized $d$ electrons
using norm-conserving pseudopotentials\cite{PSP1,PSP2}. Since there has been no theoretical 
and/or experimental determination of $U$ and $J$ for Co $d$ 
electrons in Na$_x$CoO$_2$, we adopt a moderate $U=5.5$ eV and a $J=0.9$ eV 
in our calculations and neglect their doping dependence.
The exchange parameter $J$ is of the order of
1 eV for most later transition metal oxides\cite{LDA+U1} and 
Singh has given an estimate of $U=5\sim 8$ eV for Na$_x$CoO$_2$\cite{NaCo2O4}. 
Similar values of $U$ (5.4 and 5.0 eV) have been used in previous
studies on this system\cite{Kunes03,Zuo03}.
In general, LSDA+U results are insensitive to small variation of these parameters.


\section{Results}

\subsection{Electronic structure of CoO$_2$ and CoO$_2^{1.0-}$}

We first study the undoped parents material CoO$_2$ in its ferromagnetic phase.
Single layer CoO$_2$ has a $D_{3d}$ point group symmetry, which derives
from the cubic ($O_h$) symmetry after a distortion along the [111] direction.
The Co $3d$ orbitals split into a triplet ($t_{2g}$) and a doublet ($e_g$) 
under the influence of the octahedral (cubic) crystal field.
Upon further lowering the symmetry, the $t_{2g}$ states split into $e_{g}$ 
and $a_{1g}$ levels. The $t_{2g}$ derived $e_g$ states 
then mix with the original $e_g$ ones, forming two new doublets $e_g^{(1)}$ 
and $e_g^{(2)}$. Of course, the degree of this mixing increases with
increasing trigonal distortion and Co $d$ derived $e_g$ states will further
hybridize with the O $p$ states. It is generally believed that the relevant 
low-energy electronic states of CoO$_2$ are predominately of Co $d$ character 
and can be interpreted in terms of those of the Co ion. For example, the electronic 
structure of undoped CoO$_2$ in its low spin state ($S=\frac{1}{2}$) may be 
understood in terms of Co$^{4+}$($e_{g}^{\uparrow}a_{1g}^\uparrow e_g^{\downarrow}$). 
Thus upon electron doping, it becomes a doped spin-$\frac{1}{2}$ system. However,
due to the strong mixing between O $p$ and Co $d$ states in these systems,
the validity of such a simplified picture needs to be carefully examined.

\begin{figure}[h]
\includegraphics[width=9cm]{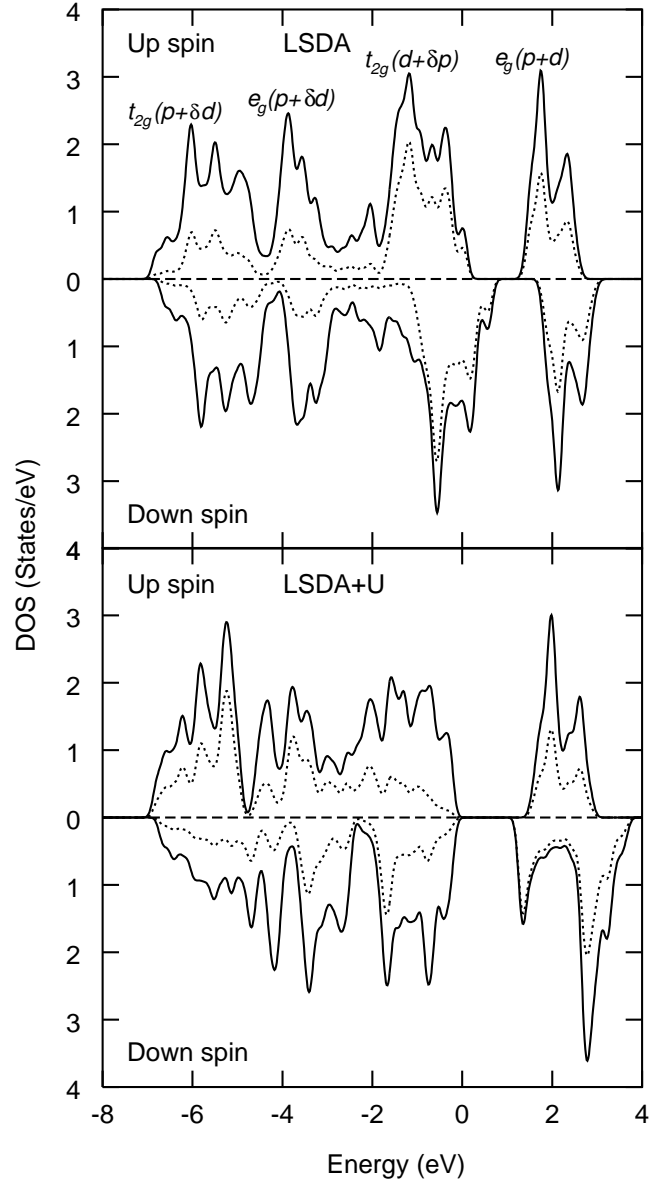}
\caption{Spin resolved DOS of CoO$_2$ in its ferromagnetic phase calculated with
LSDA (upper) and LSDA+U (lower). Solid curves show the total DOS and the dotted ones
are partial DOS projected on Co $d$ orbitals.}
\label{coo2_dos}
\end{figure}

Figure \ref{coo2_dos} compares the LSDA and LSDA+U density of states (DOS) of
CoO$_2$. The undoped parent material CoO$_2$ is a Mott-Hubbard insulator 
(or charge transfer insulator according to the ZAS classification\cite{ZAS}) 
as predicted by LSDA+U. In contrast, our LSDA calculation gives a metallic ground state 
with a rather high DOS (2.5 electrons/eV/cell) at the Fermi level,
which is consistent with the LSDA result of Singh\cite{NaCo2O4}. The local spin moment 
of Co calculated within LSDA+U is about 1 $\mu_B$, as expected for a 
spin-$\frac{1}{2}$ system. LSDA, on the other hand, gives a local moment of 
0.58 $\mu_B$ due to lack of orbital polarizations. The octahedral crystal 
field splitting ($\sim$ 3.0 eV) of Co $d$ orbitals, i.e., 
the splitting between the occupied $t_{2g}$ and the unoccupied $e_{g}$ states 
of the $d$-orbitals, can easily be estimated from the LSDA results.
Further splitting of the $t_{2g}$ states (the separation between the two peaks 
within the triplet labelled $t_{2g}(d+\delta p)$ in the upper panel of 
Fig.\ \ref{coo2_dos}) due to the trigonal distortion is about 1.0 eV within LSDA. 
Determining these values from LSDA+U results, however, is more involved
since additional splitting due to the Coulomb $U$ can not be easily decoupled.
The explicit removal of the self-interaction in the screened HF
interaction term in the LSDA+U method separates the occupied and the unoccupied $d$ states 
and pushes the occupied ones below the O $p$ levels. Consequently, the top of 
the valene band has predominately O $p$ character within LSDA+U, contrary 
to the Co $d$ character in LSDA. Another interesting observation is that 
the hybridization between the occupied Co $d$ and O $p$ 
states is enhanced within the LSDA+U method. Whereas the DOS calculated with LSDA 
clearly shows $p$-dominate $t_{2g}$ and $e_{g}$ states (labelled $t_{2g}$($p+\delta d)$ 
and $e_{g}$($p+\delta d$) in Fig.\ \ref{coo2_dos}) and $d$-dominate $t_{2g}$ 
states (labelled $t_{2g}$($d+\delta p$)), this separation between $d$ and $p$ 
states becomes less obvious in LSDA+U results. Therefore, our results suggest that both
Co $d$ and O $p$ states need to be considered when one attempts to derive an
effective low energy model hamiltonian for this system. 

\begin{figure}[h]
\includegraphics[width=9cm]{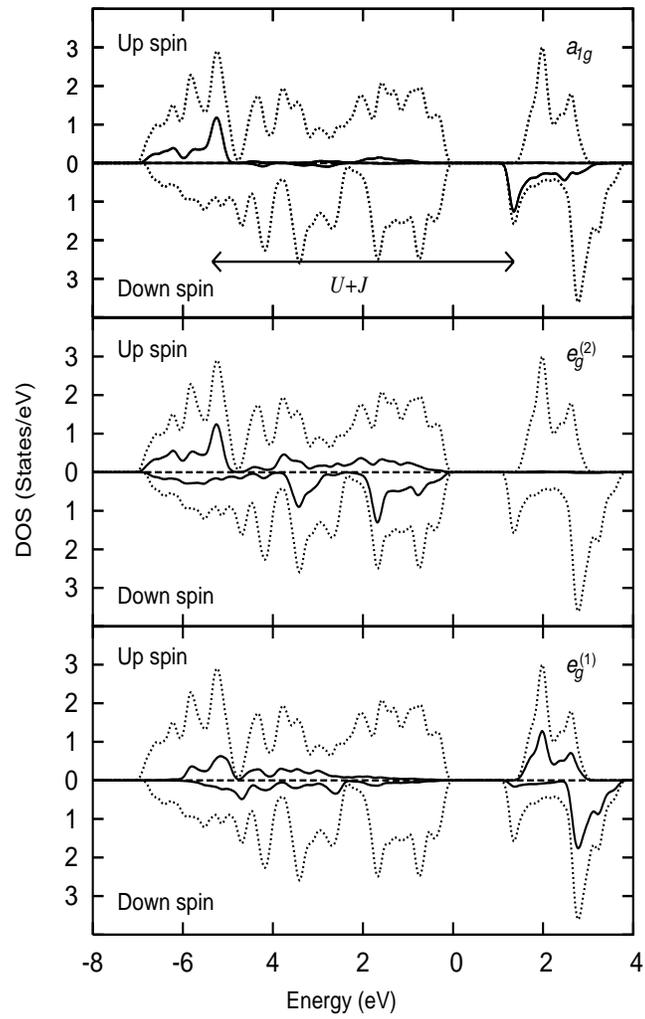}
\caption{Local DOS of CoO$_2$ projected onto the symmetry-adapted Co $d$ orbitals. The total
DOS are also shown in dotted curves.}
\label{coo2_dos2}
\end{figure}

To better understand the electronic structure of CoO$_2$, we further project the 
wavefunctions onto the symmetry-adapted Co $d$ orbitals (i.e., the eigenfunctions of 
the density matrix defined in Eq. (\ref{density_matrix})), as shown in 
Fig.\ \ref{coo2_dos2}. The corresponding orbital occupations are given in 
Table \ref{occup_coo2}. As we have mentioend above, the $t_{2g}$ triplet splits 
into $e_{g}$ and $a_{1g}$ under the influence of triangular crystal field. 
Although this splitting ($\sim$ 1.0 eV) is insignificant in LSDA, the
strong on-site Coulomb interaction included in the LSDA+U pushes the minority 
spin $a_{1g}$ up so that it becomes completely unoccupied. The separation between the 
occupied and the unoccupied $a_{1g}$ states is $U+J \sim 6.4$ eV,
as shown in the upper panel of Fig.\ \ref{coo2_dos2}. The $t_{2g}$ derived 
$e_{g}$ doublet, labelled $e_{g}^{(2)}$, is nearly fully occupied 
as expected (see Table \ref{occup_coo2}). For the other doublet, i.e., $e_{g}^{(1)}$,
one would expect it to be unoccupied from the simple molecular orbital 
analysis. However, this does not seem to be the case in our calculations due to 
strong mixing between Co $d$ and O $p$ states with compatible symmetry.
We call these hybridized doublet $e_{g}(pd)$. Both the occupied and the unoccupied  
$e_{g}(pd)$ have nearly equal O $p$ and Co $d$ characters. As a result, the 
valency of the Co ion in CoO$_2$ deviates substantially from its norminal value 4+. 
This indicates a coexistence of ionic and covalent bondings in this system: Whereas the
Co $4s$ electrons are fully ionized, the $d$ electrons are better described as
covalently bonded with O $p$ electrons due to the significant overlap between
the two wavefunctions. 

\begin{table}
\begin{center}
\begin{tabular}{cccc}
\hline
\hline
&\mbox{\hspace{1.0cm}}&\multicolumn{2}{c}{$d$ orbital occupation}\\
\multicolumn{2}{c}{}&~~~LSDA~~~&~~LSDA+U~~\\ \hline
&$a_{1g}$&0.845&0.926\\
Majority&$e_g^{(2)}$ &0.907&0.919\\
Spin&$e_g^{(1)}$ &0.457&0.506\\
&Total&3.57&3.78\\ \hline
&$a_{1g}$&0.668&0.097\\
Minority&$e_g^{(2)}$ &0.725&0.916\\
Spin&$e_g^{(1)}$   &0.434&0.383\\
&Total&2.99&2.69\\ \hline
\multicolumn{2}{c}{Total $d$ electrons}& 6.56&6.47\\
\multicolumn{2}{c}{Co Spin Moment ($\mu_B$)}& 0.58&1.09\\
\hline
\hline
\end{tabular}
\vspace{3mm}
\caption{Comparison of $d$ electron occupation of undoped CoO$_2$ between
LSDA and LSDA+U results. 
For doublets, the number of $d$ electrons is the occupation times two.}
\label{occup_coo2}
\end{center}
\end{table}

\begin{figure}[h]
\includegraphics[width=9cm]{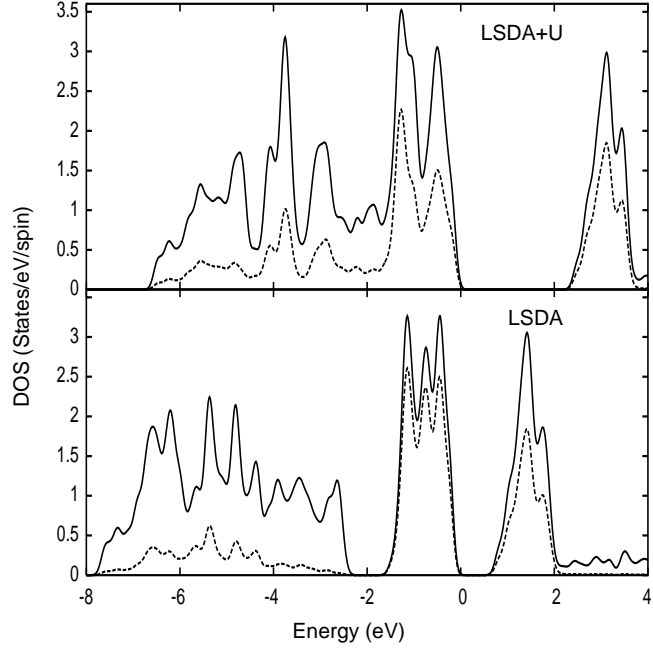}
\caption{Comparison between LSDA and LSDA+U DOS of (CoO$_2$)$^{1.0-}$. Solid curves are
total DOS and dotted ones are projections onto Co $d$ orbitals.}
\label{e1lsda_lsdau}
\end{figure}

Upon doping one electron to the CoO$_2$ layer, the system becomes
a non-magnetic insulator with a band gap of about 2.2 eV (see 
Fig.\ \ref{e1lsda_lsdau}), which compares favourably with the measured 
band gap ($\sim$ 2.7 eV) of a similar material LiCoO$_2$\cite{LiCoO2}. 
Interestingly, LSDA also predicts a non-magnetic insulating ground 
state but with a smaller gap of about 0.8 eV (see Fig.\ \ref{e1lsda_lsdau}). 
This is expected since the crystal field splitting of Co $d$ states
within LSDA is larger than the bandwidth of the $t_{2g}$ and
$e_g$ sub-bands. Therefore, NaCoO$_2$ is a band insulator within
LDA. The on-site Coulomb interaction thus contributes 
about 1.4 eV to the calculated band gap, which is only a fraction
of the parameter $U$. This is an indication that the results of LSDA+U 
calculations are not especially sensitive to the parameters.
Although both LSDA and LSDA+U give qualitatively the same nonmagnetic
insulating ground state, there are significant differences 
in the calculated DOS with the two methods, especially for the
occupied states. As for the undoped case, LSDA+U enhances the hybridization 
between the Co $d$ and O $p$ states. The top valence band triplet
has predominately Co $d$ character in the absence of local Coulomb
interactions but strongly hybridizes with O $p$ states within LSDA+U.
Also, there is a small but not negligible gap between the $d$-dominate
and $p$-dominate states within LSDA, a feature that does not exist 
within LSDA+U and was not observed in photo-emission experiments\cite{LiCoO2}.
Overall, our results show that the $d$-states mostly
concentrate within a 2 eV window below the valence band maximum and
the oxygen $p$ states spread from $-7.0$ to $-2.0$ eV. This is
in good agreement with the resonant photoemission
experiment results for LiCoO$_2$\cite{LiCoO2} where the sharp peak around
-1.4 eV is assigned to the Co $d$ final states whereas the broad
structure at -5$\pm$2 eV is attributed to O $p$ states. Of course,
one must be cautious when comparing the calculated DOS with
the photoemission results since both matrix element effects and
correlations may change the lineshape of the photoemission spectra.

The $d$ orbital occupation of both 
majority and minority spins are significantly affected by the electron 
doping (see Tables \ref{occup_coo2} and \ref{occup_coo2_1}). Whereas the 
minority spin occupation increases by 0.68 electrons within LSDA+U, the majority 
spin occupation actually decreases by 0.41. This is not a surprising
result for a strongly correlated system: Although the lower and upper Hubbard bands 
(LHB and UHB) are energetically separated, they are intimately correlated 
and are both affected by doping. Strong doping dependence
of both the band energies and the spectra weights of the correlated $d$ bands 
can also be seen by comparing the projected DOS between the undoped and 
doped cases (see the lower panel of Fig.\ \ref{coo2_dos} and the
upper panel of Fig.\ \ref{e1lsda_lsdau}). Due to their mixing
with the correlated $d$ states, O $p$ states are also affected by
the doping. 
Overall, the total $d$ electron occupation only increases by 0.27 
electrons upon doping one electron to the system. This suggests that a substantial
portion of the doped charge actually go to oxygen sites and
that the Co valency is rather insensitive to the doping level 
in this particular system. This insensitivity is also recently reported in
Na$_x$Co$_{1-y}$Mn$_{y}$O$_2$ systems, where Co ions are partially
substituted by Mn\cite{NaCoMnO2}. 

\begin{table}
\begin{center}
\begin{tabular}{ccc}
\hline
\hline
\mbox{\hspace{1.0cm}}&\multicolumn{2}{c}{$d$ orbital occupation/spin}\\
&~~~LSDA~~~&~~LSDA+U~~\\ \hline
$a_{1g}$&0.887&0.908\\
$e_g^{(2)}$&0.889&0.908\\
$e_g^{(1)}$&0.363&0.321\\
\hline
Total&3.39&3.37\\ \hline
\hline
\end{tabular}
\vspace{3mm}
\caption{$d$ electron occupation of (CoO$_2$)$^{1.0-}$ calculated
with LSDA and LSDA+U. The system is non-magnetic.
For doublets, the number of $d$ electrons is the occupation times two.}
\label{occup_coo2_1}
\end{center}
\end{table}

\begin{figure}[h]
\includegraphics[width=9cm]{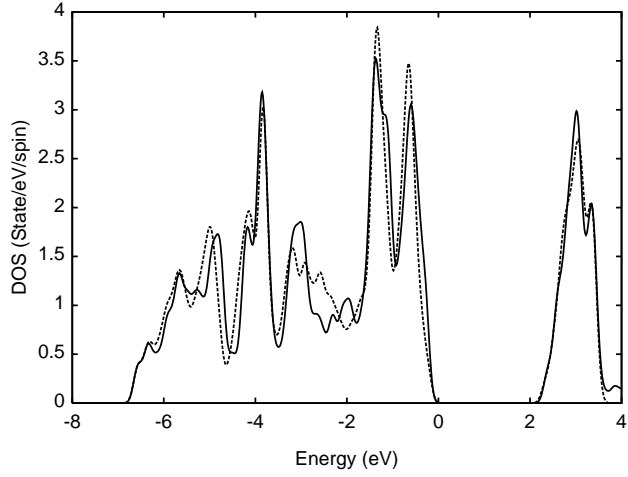}
\caption{DOS of (CoO$_2$)$^{1.0-}$ (solid curve) and that of NaCoO$_2$
(dash curve). The DOS is not altered significantly by replacing Na ions with
a uniform positive background. Also, interlayer coupling does not seem to be
important in this system.}
\label{e1nacoo2}
\end{figure}

To ensure the validity of replacing Na ions with a uniform positive 
background, which reduces the complex system to a single-layer one,
we have also done a calculation for a realistic system, i.e., a double-layer 
NaCoO$_2$. Fig.\ \ref{e1nacoo2} compares the DOS of the model
system (CoO$_2$)$^{1.0-}$ with that of NaCoO$_2$. The negligible
differences between the DOS of the two systems suggest that
our model system is appropriate and interlayer coupling
is not very important in this system, assuming no charge and/or magnetic
orderings are involved. The possibility of a subtle interplay between Na potentials
and charge orderings in the CoO$_2$ layer will be discussed later.

\subsection{Electronic structure of (CoO$_2$)$^{x-}$ ($0\leq x\leq 1$)}

\begin{figure}[h]
\includegraphics[width=9cm]{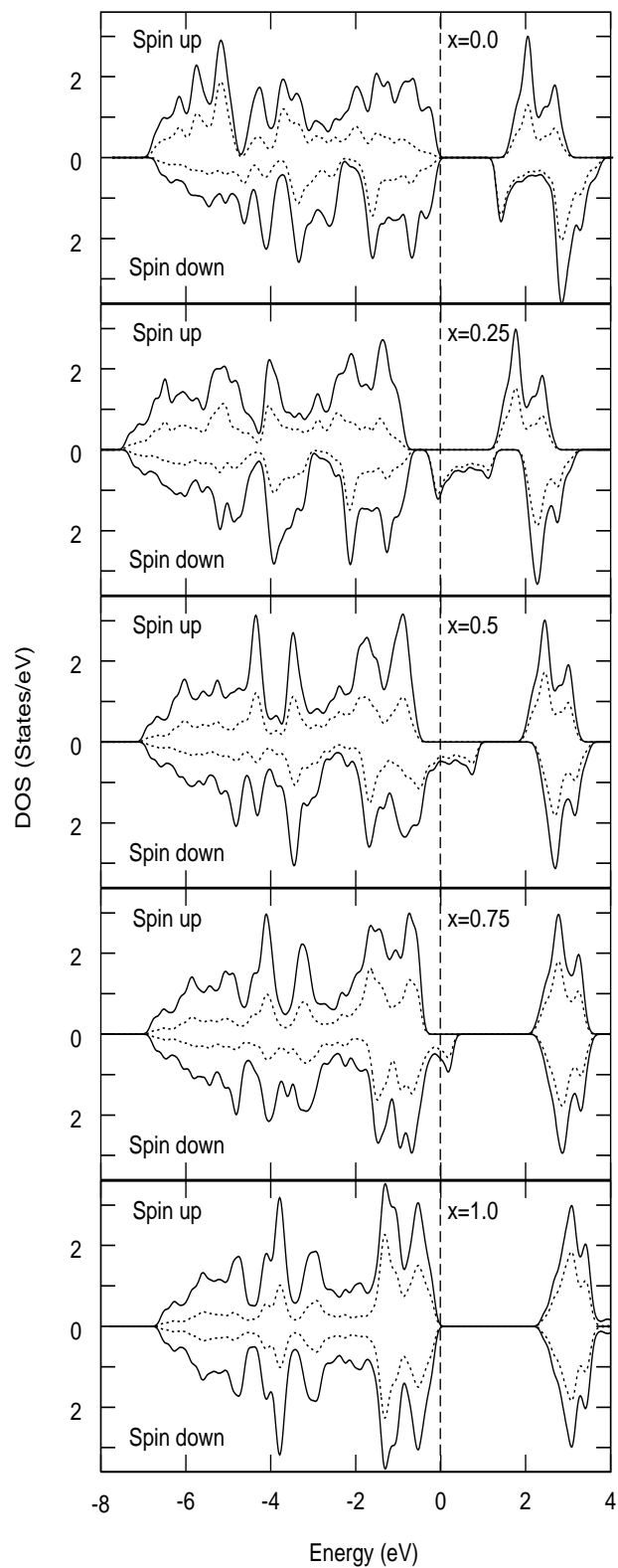}
\caption{Doping effects on the DOS of (CoO$_2$)$^{x-}$ ($0.0\leq x \leq 1.0$). 
Solid curves are total DOS and the dotted ones are projections onto
Co $d$ orbitals. As the doping level increases, the partially occupied
minority spin $a_{1g}$ state moves across the band gap while pushing other correlated
$d$ states upwards. The dashed line shows the Fermi level of metallic systems.}
\label{dos_doping}
\end{figure}

Having discussed the electronic properties of the two extreme cases,
it would be interesting to see how the electronic structure of
(CoO$_2$)$^{x-}$ evolves as the doping level $x$ varies. Fig.\ \ref{dos_doping} 
shows the evolution of the DOS as the doping level $x$ changes from 0.0 to 
1.0. As we have discussed, the undoped parent material CoO$_2$ is a charge 
transfer insulator with a gap between the O $p$ and Co $d$ ($a_{1g}$) states. 
Upon electron doping, the originally unoccupied $a_{1g}$ state, which is 
split-off from the $t_{2g}$ triplet due to on-site Coulomb interactions, 
becomes partially occupied and moves toward lower energy, touching the O 
$p$-dominate valence band at $x\sim 0.3$ and eventually merging into the 
rest of the valence band. At $x=1.0$ the $t_{2g}$ triplet is 
recovered and occupied by both spin up and spin down electrons.
At the same time, a new gap (larger than that of the undoped case) appears 
between the $t_{2g}$ and $e_g$ states. All other occupied $d$ states,
in contrast, are pushed upward as doping level increases.
As a result, significant spectra weight moves across the 
entire valence bandwidth due to the correlation between the 
doped electron and other occupied $d$ states. 

The lowering of the energy of the minority spin $a_{1g}$ 
(relative to other $d$ states) upon electon doping is intriguing and should be
studied in more detail. Table \ref{occup_doping} shows the $d$ electron occupations
for all symmetrized $d$ orbitals. 
The occupation of all other $d$ states except the minority spin $a_{1g}$ decreases
as doping level increases, with the majority spin $e_g^{(1)}$ state
being affected the most due to less hybridization between this state
and the O $p$ states as doping increases. Consequently, the effective Coulomb repulsion
``felt" by the minority spin $a_{1g}$ electron is actually
reduced with increasing doping level. Note also that the occupation of
this minority spin $a_{1g}$ state is approximately the same as the
doping level. However, this does not mean that all doped electrons
go to Co site. In fact, it seems that a significant portion of the doped
charge actually goes to oxygen sites, as we discussed previously for
the case of (CoO$_2$)$^{1.0-}$. Table \ref{occup_doping} also gives the 
local spin moment on Co site as a function of doping level, which decreases 
monotonically with increasing doping ($m_s\sim 1.0-x$ $\mu_B$). 
We should point out that 
the occupation analysis in our calculations is only approximate and
different charge analyses may give slightly different results. 
The subtle changes to the $e_g^{(2)}$ and majority spin $a_{1g}$ 
occupation might be partially due to orbital and/or structural relaxation effects.

Another factor that contributes to the doping dependence of the energy of the
minority spin $a_{1g}$ state comes from the doping dependence of 
the trigonal distortion: The crystal field splitting within the $t_{2g}$
triplet decreases with decreasing distortion. As we will show in the
following, trigonal distortion increases as doping level decreases,
and so does the splitting.
 
\begin{table}
\begin{center}
\begin{tabular}{ccccccc}
\hline
\hline
\multicolumn{2}{c}{doping}&0.0&0.25&0.5&0.75&1.0\\ \hline
&$a_{1g}$&           0.926&0.922&0.917&0.912&0.908\\
Majority&$e_g^{(2)}$&    0.919&0.916&0.914&0.911&0.908\\ 
Spin&$e_g^{(1)}$&0.506&0.436&0.381&0.345&0.321\\
\cline{2-7}
&total&              ~3.78~~&~3.63~~&~3.51~~&~3.42~~&~3.37~~\\
\hline
&$a_{1g}$&           0.097&0.341&0.566&0.754&0.908\\
Minority&$e_g^{(2)}$&    0.916&0.913&0.910&0.908&0.908\\ 
Spin&$e_g^{(1)}$&0.383&0.379&0.367&0.346&0.321\\
\cline{2-7}
&total&              2.69 &2.93 &3.12 &3.26 &3.37\\
\hline
\multicolumn{2}{c}{Co Spin Moment ($\mu_B$)}&1.08&0.70&0.39&0.16&0.0\\ \hline
\hline
\end{tabular}
\vspace{3mm}
\caption{$d$ electron occupation and local spin moment of (CoO$_2$)$^{x-}$ as a function of doping level.
For doublets, the number of $d$ electrons is the occupation times two.}
\label{occup_doping}
\end{center}
\end{table}

\begin{figure}[h]
\includegraphics[width=9cm]{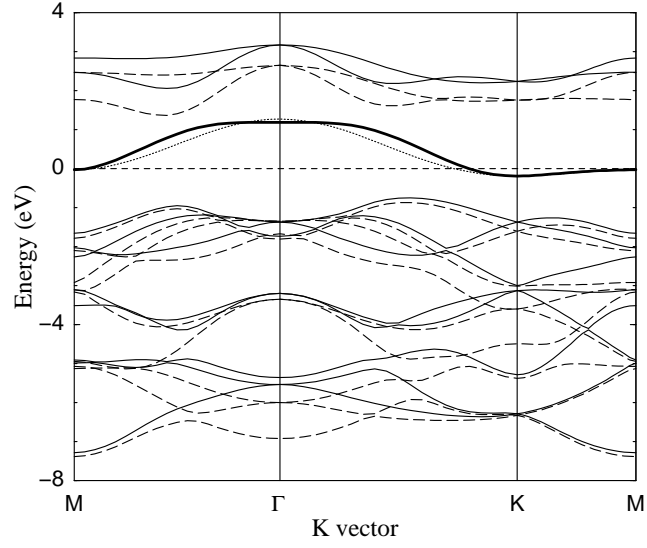}
\caption{Band structure of (CoO$_2$)$^{0.25-}$. Solid and dash curves are for minority
and majority spins, respectively. The minority spin $a_{1g}$ band
is highlighted with a thick solid curve and the dotted curve is a nearest 
neighbor tight-binding fitting for this band with a hopping parameter $t=0.155$ eV.}
\label{band_025}
\end{figure}

\begin{figure}[h]
\includegraphics[width=9cm]{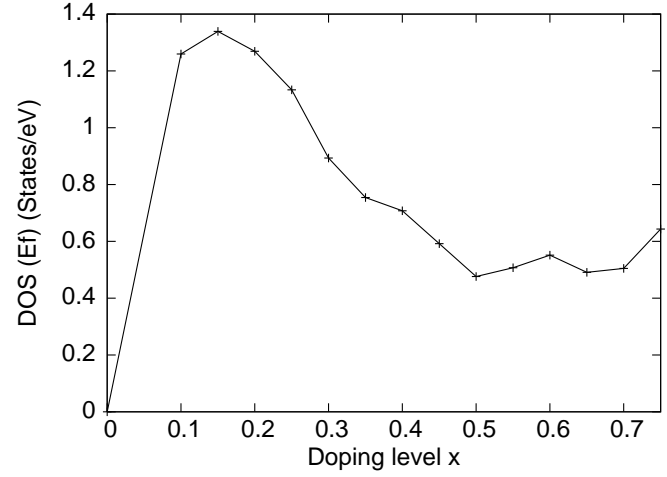}
\caption{Density of states of (CoO$_2$)$^{1.0-}$ at the Fermi level 
as a function of doping level $x$.}
\label{dos_ef_doping}
\end{figure}

The minority spin $a_{1g}$ state deserves particular attention
since, upon electron doping, it determines the low-energy electronic
properties of the system and has been the subject of intensive
discussion concerning its connection with the superconductivity observed
in Na$_{0.3}$CoO$_2$$\cdot$yH$_2$O\cite{Baskaran03,Tanaka03,Honerkamp03,Wang03,Chou03}.
Fig.\ \ref{band_025} shows the band structure of (CoO$_2$)$^{0.25-}$ with
the minority spin $a_{1g}$ band highlighted. Interestingly, apart from
a constant shift, this band can be fairly well fitted by a simple
tight-binding model with a nearest hopping parameter $t=-0.155$ eV.
Note that here $t$ is not the hopping element between Co and O 
sites but the effective hopping between neighboring Co sites.
The total bandwidth is thus $9|t|=1.4$ eV. Since we use a on-site Coulomb 
interaction $U=5.5$ eV, the effective superexchange between two neighboring 
Co ions is $J=4t^2/U=16$ meV, in reasonable agreement with a previous
estimate\cite{Wang03}. Due to the particular dispersion and doping
dependence of this band,
there is a strong doping dependence to the Fermi-surface properties and 
the DOS at the Fermi level, as shown in Fig.\ \ref{dos_ef_doping}. The DOS($E_f$) 
increases sharply with increasing doping level for $x\leq 0.1$, 
reaches a maximum at about $x\sim 0.2$, and then decreases with 
increasing doping. The narrow doping range $0.1\leq x\leq 0.3$ 
beyond which DOS($E_f$) decreases rapidly with increasing or decreasing 
doping level is closely related to the Fermi surface structure of the 
system: At very low dopings, small electron pockets appear around the corners
of the BZ. (Note that this metallic state may not be stable against charge
orderings at very low doping levels.) The Fermi surface quickly extends  
and then shrinks with increasing doping. Compared 
with LSDA results\cite{NaCo2O4} where a large Fermi surface, as well as small 
pockets of holes, are predicted, there is only one large Fermi surface in our 
calculation, which agrees well with a recent experiment\cite{Fermi_Surface}.
Although it would be interesting to connect this observation with the fact 
that superconductivity occurs only in a very narrow doping range ($x\sim 0.3$ in 
Na$_{x}$CoO$_2$$\cdot$yH$_2$O\cite{supercond,supercond2}), further investigation 
on this subject is required. If the superconductivity in this system is of phonon 
origin, then the high DOS at the Fermi level is definitely an important factor 
in determining the superconducting transition temperature. 

\subsection{Doping effects on the structural properties and possible spin-phonon interactions}

\begin{figure}[h]
\includegraphics[width=9cm]{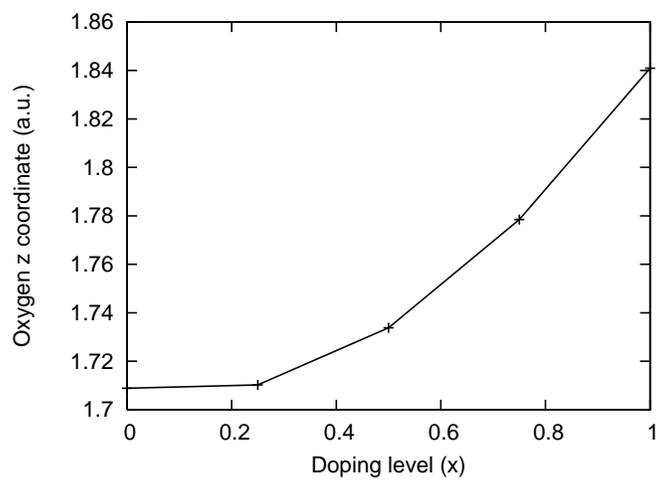}
\caption{Calculated equilibrium  oxygen $z$ coordinate of (CoO$_2$)$^{x-}$ 
as a function of doping level $x$.}
\label{oz_doping}
\end{figure}

Structural properties are in general not particularly sensitive to the
doping level. As we have mentioned above, we fix the lattice constants
in our calculations but allow the oxygen atoms to relax. Fig.\ \ref{oz_doping} shows 
the doping-dependent oxygen $z$ coordinate as measured from the Co plane.
The calculated O $z$ coordinate at $x\sim 0.30$ (1.72 a.u.) falls within the measured 
values (1.67 $\sim$ 1.77 a.u.) for Na$_x$CoO$_2\cdot y$H$_2$O\cite{supercond,naxcoo2_structure}
but is smaller than those ($\sim$ 1.83 a.u.) for unhydrated systems\cite{naxcoo2_structure}.
This is reasonable since the single layer system in our calculations should mimic the
hydrated system better than the unhydrated one. The Na potential in the
unhydrated system is likely to attract the negatively charged O ions away from the Co
layer, making the O $z$ coordinate larger. Overall, the distance between the oxygen and 
cobalt layers expands quadratically with increasing doping level. 

Using the LSDA+U total energy functional defined in Eq. (\ref{ELSDA+U}), we can
calculate phonon energies under the frozen phonon approximation\cite{Yin82}. 
Single layer CoO$_2$ (assuming paramagnetic/ferromagnetic ordering) has four 
zone-center optical phonon modes. Two of them relate to the in-plane and out-of-plane 
motion of oxygen atoms ($E_{g}$ and $A_{1g}$), which are Raman active. The other two 
($E_{u}$ and $A_{2u}$) involve cobalt moving against oxygen and are infrared (IR) active. 
Table \ref{phonon} lists the calculated zone center phonon energies for doping levels 
$x=$ 0.0, 0.25, 0.5, 0.75 and 1.0. The calculated energies are in good agreement with 
available measurements\cite{raman1,raman2,IR}. For example, the measured $A_{1g}$ phonon 
energy (ranging from 71.2 to 74.4 meV\cite{raman1,raman2} depending on the doping level 
and sample conditions) agrees well with the calculated ones (from 72.6 to 73.0 
meV for doping level $0.25\leq x\leq 0.75$). The theoretical $E_{g}$ phonon energies
are 63.2 meV for $x=0.25$ and 65.6 meV for $x=0.75$, to be compared with the measured 
values 56.8 $\sim$ 61.2 meV\cite{raman1}. The measured energy for the IR active mode 
$E_u$ is about 70.7 
meV for Na$_{0.57}$CoO$_2$\cite{IR}, which also compares favorably with our theoretical
value (74.7 meV for $x=0.5$). There has been no measurement for the $A_{2u}$ mode so far. 
In general, the zone center optical phonons are not sensitive to the doping level. There is, 
however, one interesting exception: At $\sim$ 0.25 doping, the $E_g$ phonon softens, 
decreasing from 67.8 meV for $x=0.0$ to 63.2 meV for $x=0.25$, and has significant 
anharmonicity. (The calculated harmonic phonon energy is only 56.5 meV.) The softening of
this phonon mode at doping level $x\leq 0.2$ seems to indicate a strong electron-phonon
coupling and we believe that the high DOS at the Fermi level for $x\sim 0.25$, 
together with strong electron-phonon couplings between this mode and 
conducting states, is responsible for the phonon softening and anharmonicity and
may eventually lead to a superconducting phase transition.  

\begin{figure}[h]
\includegraphics[width=8.5cm]{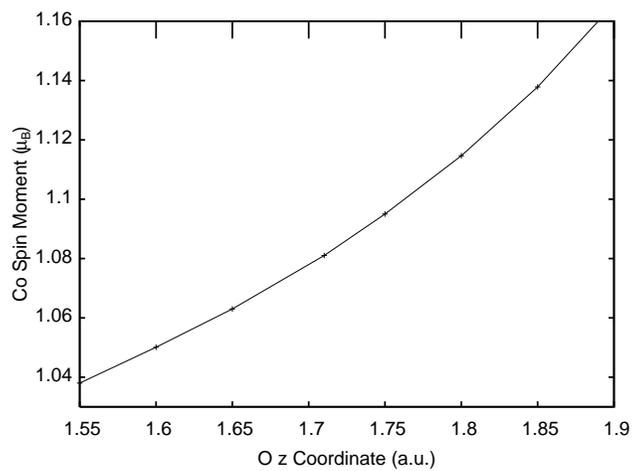}
\caption{Correlation between the local spin moment on Co and the oxygen $z$ coordinate of
CoO$_2$.}
\label{oz_spin}
\end{figure}

In magnetic systems, phonons might interact with the spin degree of freedom.
In fact, we observe a strong correlation between the O $z$ coordinate (relates to
the $A_{1g}$ phonon displacement) and the local spin moment on Co sites (see 
Fig.\ \ref{oz_spin}). No such correlations were found for other phonon modes. This raises 
the possibility of interactions between $A_{1g}$ phonons and magnons in this
system.

\begin{table}
\begin{center}
\begin{tabular}{cccccc}
\hline
\hline
&\multicolumn{5}{c}{zone center phonon energy (meV)}\\ 
doping level&0.0&~~~~0.25~~~~&0.5&0.75&1.0\\ \hline
$A_{1g}$&72.3&72.6&73.4&73.0&71.4\\
$A_{2u}$&72.8&73.6&74.0&77.0&80.0\\
$E_{g}$&67.8&63.2 (56.5)&64.6&65.2&62.7\\
$E_{u}$&75.3&74.9&74.7&74.2&72.7\\
\hline
\hline
\end{tabular}
\vspace{3mm}
\caption{Doping effects on zone center phonon energies calculated 
using frozen phonon approximation. All phonon modes are fairly harmonic
except the $E_g$ mode in the case of doping level $x \sim 0.25$. For
$x=0.25$, the calculated harmonic frequency is shown in parentheses.}
\label{phonon}
\end{center}
\end{table}

\section{Discussions}

\subsection{Other possible spin configurations}

So far we have assumed a low spin state for Co ions and the doped electrons always 
go to the minority spin $a_{1g}$ conduction band. Thus the local spin moment on Co 
decreases with increasing doping level and vanishes at $x=1.0$. However, this does 
not seem to be consistent with the observations that sizable effective magnetic moments 
$\mu_{\mathrm{eff}}$ exist at doping level $x\sim 0.75$. For example, magnetic susceptibility 
measurements of Na$_{0.75}$CoO$_2$ give a $\mu_{\mathrm{eff}}$ of 2.74 $\mu_B$/Co assuming 
only Co$^{4+}$ ions contribute to the Curie constant $C$\cite{magnetic_transition1}. 
This is much larger than the ``spin-only" value ($g\sqrt{s(s+1)}=\sqrt{3}\mu_B$) of 
Co$^{4+}$ ion in its low spin state. If, on the other hand, all Co sites are assumed 
equivalent, the results yield a $\mu_{\mathrm{eff}}$ of 1.37 
$\mu_B$/Co\cite{magnetic_transition1}, 
which is again not compatible with the calculated spin moment.  
This apparent discrepancy raises the possibilities of unquenched orbital moments 
and/or other spin configurations of Co ions in this system, especially for high 
doping levels ($x\ge 0.5$). The spin states of Co ions in CoO$_2$ are determined 
by several competing factors such as the crystal field splitting, Hund's 
rule coupling and the screened on-site Coulomb interactions. Therefore, depending 
on the relative strength of these factors, some Co ions might adopt an intermediate-spin 
state. Here we explore such a possibility.

As we can see from Fig.\ \ref{coo2_dos}, the energy of the unoccupied majority 
spin $e_g$ is only slightly higher than that of minority spin $a_{1g}$ in the 
undoped single layer CoO$_2$. In the real system, especially in unhydrated 
Na$_x$CoO$_2$, however, the situation might be more complicated. The presence of 
a possibly ordered Na potential might enhance or induce charge orderings in the 
CoO$_2$ layer\cite{Foo03} and the energy level of the two unoccupied states, 
$e_g$ and $a_{1g}$, might get reversed on some Co sites.
A smaller crystal field splitting or larger $U$ could pull down the $e_g$ state 
or push up the $a_{1g}$ one. The doped electron could then go to the $e_g$ majority 
spin state, resulting in an intermediate-spin state for the corresponding Co ions. 
Therefore, we propose the following scenario for unhydrated Na$_x$CoO$_2$
system: At low doping level, all Co ions are in their low spin states. 
As doping level increases, Na ions tend to order themselves to minimize 
the Coulomb energy. (We will discussed Na orderings in more detail in the 
following.) This ordering might then enhance or induce charge ordering
in CoO$_2$ layers. Due to charge ordering in the CoO$_2$ layer, Co ions 
then have a different symmetry and chemical environment, leading to different 
crystal field splitting and/or on-site interactions. Under certain circumstance,
the unoccupied majority spin $e_g$ state might be lower in energy than the $a_{1g}$ state.
The local moment of these Co ions will then increase with increasing doping.
We have calculated a ferromagnetic system with Co ions in their intermediate 
spin states for doping level $x=0.5$ and found that the local spin moment is 
1.52 $\mu_B$/Co. Not surprisingly, the energy of the intermediate spin state is 
higher than that of the low-spin state (by $\sim$ 0.5 eV/Co). However, charge 
orderings might reduce or inverse this energy difference, resulting in an 
intermediate-spin ground state. It is also plausible that this kind of rich 
degeneracy of spin states, a result of nearly perfect balance between the 
crystal field and Hund's rules effects, is responsible for the unusually high 
thermopower in this system.

\subsection{Na ordering, magnetic and/or charge orderings}

It is usually assumed that the Na layer is disordered in Na$_x$CoO$_2$ and
the primary effect of Na is to provide electrons to the CoO$_2$ layer. In 
strongly correlated systems, however, charge and/or spin orderings usually 
happen at an extremely low energy scale and seemingly insignificant interactions 
can sometimes result in profound changes in the electronic structure. 
Although we have shown that the Na potential has minimal 
effects on the calculated electronic structure of ferromagnetically
ordered CoO$_2$, it is still unclear the exact role Na plays 
in determining the properties of the system, especially at high doping when 
the Na layer is likely ordered. Whether or not Na becomes ordered 
depends on a competition between entropic and energetic factors. If 
energetics dominate, an ordering might occur. In fact, there has
been increasing experimental evidence that the Na layer might be ordered 
at some doping levels. Foo {\it et al.}\ observed a $\sqrt{3}\times 2$ 
ordering in Na$_{0.5}$CoO$_2$ and discussed the possible effects of Na 
ordering on its electronic properties\cite{Foo03}. Shi {\it et al.}\ reported 
a $2\times 1$ superstructure in Na$_x$CoO$_2$ for $x\ge 0.75$ and
ascribed it to a possible Na ordering\cite{Shi04}. 

\begin{figure}[h]
\includegraphics[width=7cm]{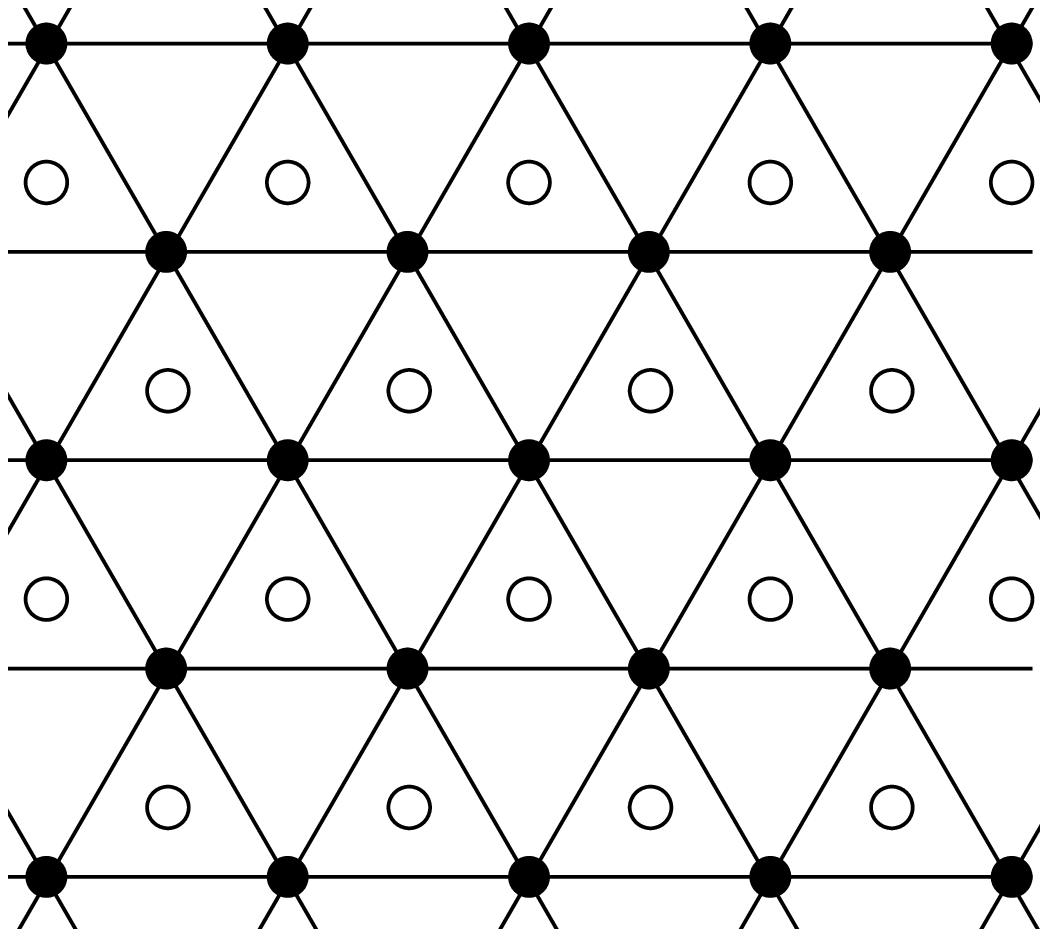}

\vspace{5mm}
\includegraphics[width=7cm]{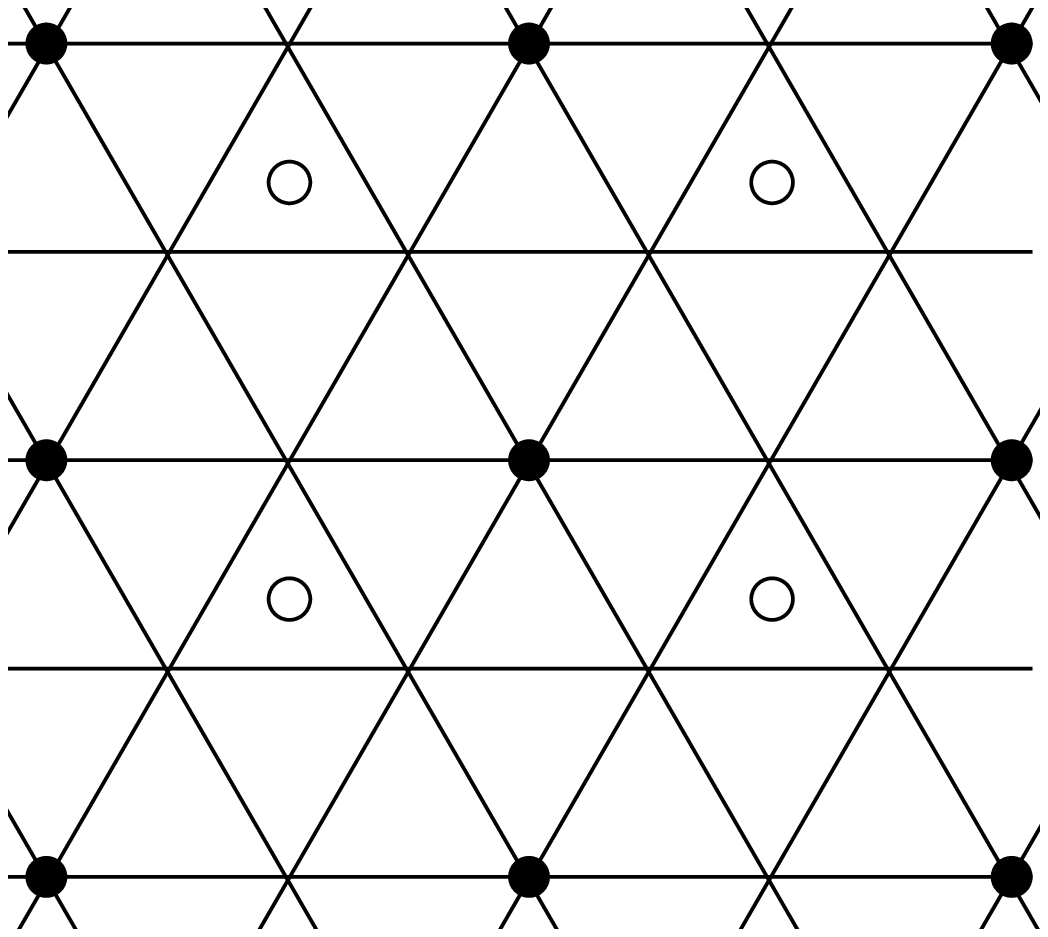}
\vspace{5mm}

\includegraphics[width=7cm]{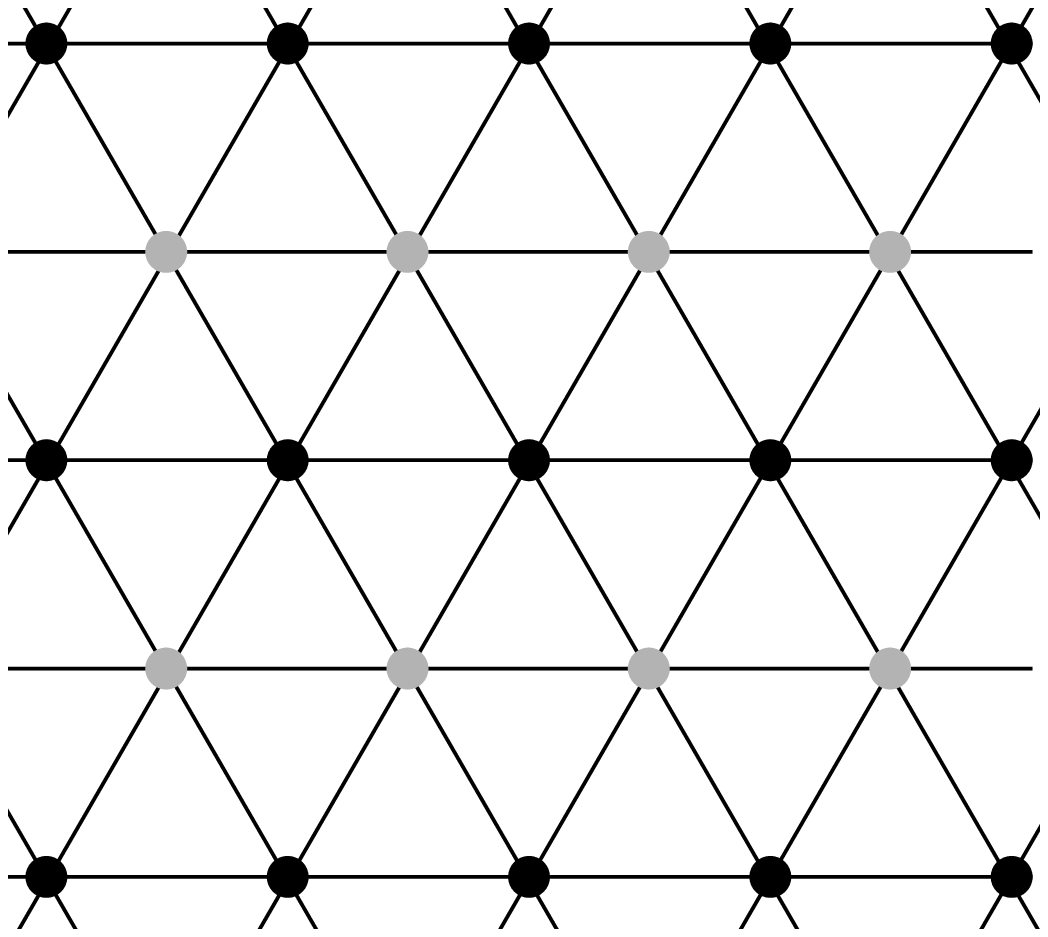}
\caption{Possible Na ordering pattens. Upper: Two distinct Na sites
in Na$_x$CoO$_2$. At doping $x=1.0$, Na will occupy only one of
the two sites at low temperature. Middle: Lowest energy ordering pattern for $x=0.5$.
Lower: A low energy ordering pattern for $x=0.75$. Gray circles
indicate partially (50\%) occupied Na sites.}
\label{na_ordering}
\end{figure}

Since Na are fully ionized, it might be possible to discuss their 
ordering by simple electrostatic and entropic arguments. (Chemical 
interactions between Na and CoO$_2$ layers might also play a minor role
but more difficult to characterize.) At low Na concentrations, 
entropic effects should dominate and the Na layer is likely disordered. 
As doping level increases, however, Na ions will tend to organize 
themselves (at least partially and locally) to minize the ionic repulsion, 
since it costs energy to place two Na ions in neighboring $2b$ and $2d$ sites 
(see Fig.\ \ref{na_ordering}). This does not necessarily lead to a long-range
ordering since there could be many (nearly) degenerate local 
orderings. However, if there exist an ordered pattern which has
significantly lower energy (compared to the thermal energy) than others, 
a long range ordering might result. For $x=0.5$, we indeed find a particular 
arrangement of Na ions which is compatible with the observed $\sqrt{3}\times 2$ 
ordering\cite{Foo03} and has as much as 0.3 eV/Na lower Coulomb energy than other
configurations with similar unit cells (see Fig.\ \ref{na_ordering}) if only in-plane 
Coulomb interactions are taken into account. It might be possible that the interplay 
between the ordering in the Na layer and the charge ordering in the CoO$_2$ layer is 
responsible for the observed insulating behavior in Na$_{0.5}$CoO$_2$, as discussed
by Foo {\it el al}\cite{Foo03}. For $x=0.75$, the situation is more complicated. We 
find many possible orderings with similar energy. The lowest-energy pattern 
is show in Fig.\ \ref{na_ordering}, with filled black circles denoting occupied  
and gray circles partially occupied (50\% for $x=0.75$) Na sites. Interestingly, 
this pattern of low energy ordering is also consistent with the reported Na superstructure
for $x\ge 0.75$\cite{Shi04}. The low energy ordering pattern for $x=0.5$
is rather exclusive in the sense that further addition of Na to this structure
will result in occupation of neighboring $2b$ and $2d$ sites thus
increase the Coulomb energy sharply. 

In general, we find it very energetically unfavorable to occupy 
neighboring $2b$ and $2d$ sites. At high doping (e.g., $x\ge$ 0.75), Na ions tend to 
occupy only one of the two distinct sites within a given domain. The size of these 
domains presumably increases with decreasing temperature. If interactions between
Na and CoO$_2$ layers are taken into account, the two Na sites may not be equilvalent
energetically. In fact, our calculations indicate that the $2d$ site has about 0.1 eV/Na 
lower in energy. This is due to the different Coulomb repulsion between the Na and Co in the 
two configurations. Therefore, $2d$ sites are more likely to be occupied, provided that
no immediate neighboring $2b$ sites are already occupied. Although this differentiation
between the two sites should be taken into account when discussing Na orderings, no 
changes to our conclusion for a Na ordering of $x=0.5$ is expected due to
the large in-plane ordering energy.

\begin{figure}[h]
\includegraphics[width=9cm]{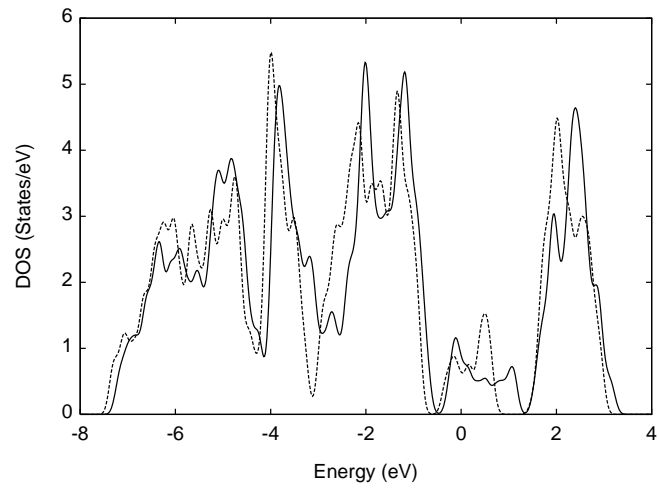}
\caption{Magnetic ordering effects on the electronic structure of
(CoO$_2$)$^{0.3-}$. The solid curve is for FM phase and the dotted curve for
AFM phase. Although no significant changes to low-lying valence 
states are observed, the width of the $a_{1g}$ conduction band is
renormalized (narrowed) appreciately due to the ordering.}
\label{e03_fm_afm}
\end{figure}

Magnetic and/or charge orderings in Na$_x$CoO$_2$ are other subjects of great interest. 
Kunes {\it et al.}\cite{Kunes03}, and Motrunich {\it at al.}\cite{Motrunich03},
discussed a possible $\sqrt{3}\times\sqrt{3}$ charge ordering for doping 
$x\sim\frac{1}{3}$. Foo {\it et al.} reported a $\sqrt{3}\times 2$ Na ordering at $x=0.5$ 
as we have mentioned above. NMR measurements also point to possible charge orderings
in Na$_x$CoO$_2$ for $0.5\le x\le 0.75$\cite{Gavilano03,Mukhamedshin04}. However, there 
is no consensus in the literature on this matter so far. The magnetic ordering in Na$_x$CoO$_2$ 
is even more intriguing. Very weak magnetic ordering has been observed at $T\sim$ 22 K 
only for doping $x\sim0.75$. Curiously enough, the measured magnetizations differ by two 
order of magnitute between two experiments\cite{magnetic_transition1,magnetic_transition2}. 
We try to explore the simplest AFM ordering with a $2\times 1$ unit cell in this system. 
The AFM phase is found to be slightly lower in energy ($\sim$ 10 meV/Co) than the FM phase 
for doping level $x=0.3$. However, this small difference could be beyond the accuracy of 
our theoretical treatment. Fig.\ \ref{e03_fm_afm} compares the calculated DOS of FM and AFM phases 
of (CoO$_2$)$^{0.3-}$. Although the low energy valence states are not significantly 
affected, the bandwidth of the partially occupied $a_{1g}$ state is greatly reduced as 
a result of AFM ordering. This raises the possibility of further supression of the 
$a_{1g}$ bandwidth if more complicated orderings are present, which might
help to account for the mysterious electron mass enhancement\cite{Chou03} in this system.
Unfortunately, due to the extremely small energy differences between competing ordering 
states, fluctuations among these states result in very slow convergence in self-consistent 
calculations. 

\section{Conclusion}

In conclusion, we have carried out systematic studies on the electronic, magnetic and 
structural properties of single layer (CoO$_2$)$^{x-}$ using a recently implemented 
rotationally invariant LSDA+U method within the pseudopotential plane-wave formalism.
Both the undoped and one integer electron doped systems are insulators within LSDA+U, 
whereas systems with fraction doping are half-metal in the absence of charge ordering
and assuming a ferromagnetic phase.
Calculated Fermi surface and zone center phonon energies agree well with available
measurements. Possible intermediate spin configurations of Co ions, Na orderings, and 
magnetic and charge orderings in this system are also discussed. 

Although the pairing mechanism that leads to superconditivity in this system remains a 
subject of intensive investigation, high DOS at the Fermi level at
low doping levels, together with strong electron-phonon couplings, might 
be partially responsible for the superconducting transition in the hydrated systems. 
The role water molecules play in the superconducting transition is still unknown. One 
possibility is that the screening effects, which greatly supress the interaction between 
Na and CoO$_2$ layers and possible charge orderings in the CoO$_2$ layer, lead to a more 
homogeneous electronic system in the CoO$_2$ layer and ultimately favor a superconducting 
state over competing phases. A better understanding of the properties of this material 
requires more experimental work and thorough theoretical investigations. The interplay 
between the Na ordering and the charge/magnetic ordering in the CoO$_2$ layer deserves 
particular attention, especially in the unhydrated system.

\begin{acknowledgments}
This work was supported by National Science Foundation Grant No. DMR-0087088,
Grant No. DMR-0213623 through the Center of Materials Simulation and Office 
of Energy Research, Office of Basic Energy Sciences, 
Materials Sciences Division of the U.S. Department of Energy under Contract 
No. DE-AC03-76SF00098. Computational Resources were provided by NPACI, MRSEC and NERSC.
\end{acknowledgments}

\newpage

\newpage

\newpage

\newpage

\newpage

\newpage

\newpage

\newpage

\newpage
\newpage

\newpage

\end{document}